\documentclass[preprint,showpacs,amsmath]{revtex4}
\usepackage{graphicx}
\usepackage{dcolumn}
\usepackage{bm}

\def\s#1{\slash\!\!\!{#1}}
\def\bee{\begin{eqnarray}}
\def\eee{\end{eqnarray}}
\def\nn{\nonumber\\}

\def\Tr{\makebox{Tr}}

\begin{document}
\title{Three Parton Corrections in $B\to PP$ decays}
\author{Tsung-Wen Yeh}
\email{twyeh@ms3.ntcu.edu.tw}
\affiliation{Department of Science Application And Dissemination, National Taichung University,
Taichung 403, Taiwan}
\begin{abstract}
The $1/m_b$ corrections from the three parton $q\bar{q}g$ Fock state of
the final state light meson in $B\to PP$ decays are evaluated by means of a collinear expansion method. 
The impacts of these corrections on the $CP$ averaged branching ratios of the $B\to \pi K$ decays are analyzed. 
\end{abstract}
\pacs{13.25.Hw \\ Keywords: power corrections, QCD factorization, B decays}

\maketitle
\section{introduction}
The QCD factorization \cite{Beneke:1999br,Beneke:2000ry,Beneke:2001ev} has been widely used to investigate 
the charmless hadronic $B$ decays.
For an operator $O_i$ of the weak effective Hamiltonian,
the matrix element for $\bar{B}\to M_1 M_2$ decays under the QCD factorization is found to be expressible as
\bee\label{fact-1}
\langle M_1 M_2|O_i|\bar{B}\rangle
&=&\sum_{j} F^{B\to M_1}_{j}(m_2^2)\int_0^1 du T^{I}_{ij}(u)\Phi_{M_2}(u)+(M_1\leftrightarrow M_2)\nn
&&+\int_0^1 d\xi du dv T^{II}_{i}(\xi,u,v)\Phi_{B}(\xi)\Phi_{M_1}(u)\Phi_{M_2}(v)\;,
\eee
where $T^{I(II)}$ denote the parton amplitudes 
and $\Phi_{B}$, $\Phi_{M_1}$, $\Phi_{M_2}$ represent the light-cone distribution amplitudes (LCDAs) for the 
initial state $\bar{B}$ meson and the final state $M_1$ and $M_2$ mesons, respectively.
The parton amplitudes $T^{I(II)}$ contain short distance interactions 
involved in the decay processes.
The LCDAs $\Phi_{B}$, $\Phi_{M_1}$, and $\Phi_{M_2}$ are introduced to account for the long distance interactions.
The $F^{B\to M_1}_{j}(m_2^2)$ with $j=+, 0$ are the $B\to M_1$ transition form factors.
The meson state vector $|M_{i}\rangle$, $i=1,2$, 
for the meson $M_i$ is composed of Fock states with different number of partons
\bee
|M_i\rangle=|q\bar{q}\rangle_{M_i}+|q\bar{q}g\rangle_{M_i}+\cdots\;.
\eee
So far, most applications of the factorization formula Eq.~(\ref{fact-1}) are limited to 
leading Fock state $|q\bar{q}\rangle_{M_{i}}$ of the light mesons.
However, the three parton Fock state $|q\bar{q}g\rangle_{M_{i}}$ of the $M_i$ meson can also contribute.
 
The corrections related to the higher Fock state are usually classified as subleading twist contributions, 
since their contributions are suppressed by factors of $O(1/m_b^n)$ with $n\ge 1$ 
in comparison with the leading ones. 
Here, $m_b$ denotes the $b$ quark mass.
Within QCD factorization, 
one can employ the Feynman-diagram approach  
or the effective-theory approach for studies of subleading twist contributions.
There exist established Feynman-diagram approaches for processes other than hadronic decays, 
such as the calculation scheme for the inclusive hard scattering processes \cite{Ellis:1982wd,Ellis:1982cd,Qiu:1988dn} 
or the method for the exclusive hard scattering processes 
\cite{Yeh:2002up,Yeh:2002rd,Yeh:2001ys,Yeh:2001ta,Yeh:2001gu}.
However, a systematic Feynman-diagram approach for charmless hadronic $B$ decays is still inaccessible.
On the other hand, the effective-theory approaches for charmless hadronic $B$ decays
have been extensively investigated in recent years 
\cite{Bauer:2000yr,Beneke:2002ph,Beneke:2002ni,Chay:2002vy,Hill:2002vw,Becher:2003qh}.

In this paper, a calculation scheme based on the Feynman-diagram approach  
for charmless hadronic $B$ decays will be developed.
We will concentrate on the construction of this calculation scheme 
and apply the constructed method to calculate the tree level three parton corrections.
The $O(\alpha_s)$ three parton corrections is also desirable to understand
their factorization properties.  
Since the related analysis is tedious,
we plan to present the relevant calculations in our another preparing work \cite{Yeh:2007}.
The organization of this paper is as following.
In Section II, the calculation scheme will be constructed.
The factorization of the tree level three parton corrections into the partonic and hadronic parts will be outlined. 
In Section III, the analysis on how the tree level three parton corrections 
can be factorized into its partonic and hadronic parts will be described in detail.
In Section IV, we will apply the results of the Section III to make predictions for the branching ratios of
$B\to\pi K$ decays.
The last section devotes for discussion and conclusion.

\section{collinear expansion at tree level}
In this section, we will generalize the collinear expansion method \cite{Ellis:1982wd,Ellis:1982cd,Qiu:1988dn} 
to calculate the three parton corrections from the
Fock state $|q\bar{q}g\rangle$ of the meson $M_2$ in the decay $\bar{B}\to M_1 M_2$.
There exist other types of power corrections, such as the power corrections from soft gluons or renormalons.
We identify these as non-partonic power corrections.
For these power corrections, our proposed scheme may not be useful.
However, to include these non-partonic power corrections requires further assumptions beyond the factorization.
For example, the soft gluon power corrections are better determined by nonperturbative theories, 
such as the QCD sum rules or lattice QCD.
In this work, we only investigate how the partonic (or the dynamic) power corrections can be included into
the QCD factorization in a consistent way.

The collinear expansion method arises from a motivation of generalizing the leading twist factorization theorem  
for the hard scattering processes to include the corrections from high Fock states of the target hadrons.
The original idea of the collinear expansion method was proposed by Polizer at 1980 \cite{Politzer:1980me}. 
The systematical method was developed by Ellis, Furmanski and Petronzio (EFP) 
in their pioneer works \cite{Ellis:1982wd,Ellis:1982cd}. 
Using the collinear expansion, the EFP group showed that, for the DIS processes, 
the twist-4 power suppressed corrections can be factorized into short distance and long distance parts,
which are in a similar factorized form as the leading twist contributions.
However, in the EFP's approach, the parton interpretation for the twist-4 corrections are lost. 
To recover the parton model picture, 
Qiu then introduced a Feynman-diagram approach \cite{Qiu:1988dn} to re-formula the EFP's method.
In this Feynman-diagram language, a parton model interpretation for the twist-4 corrections becomes trivial.

\begin{figure*}
\includegraphics{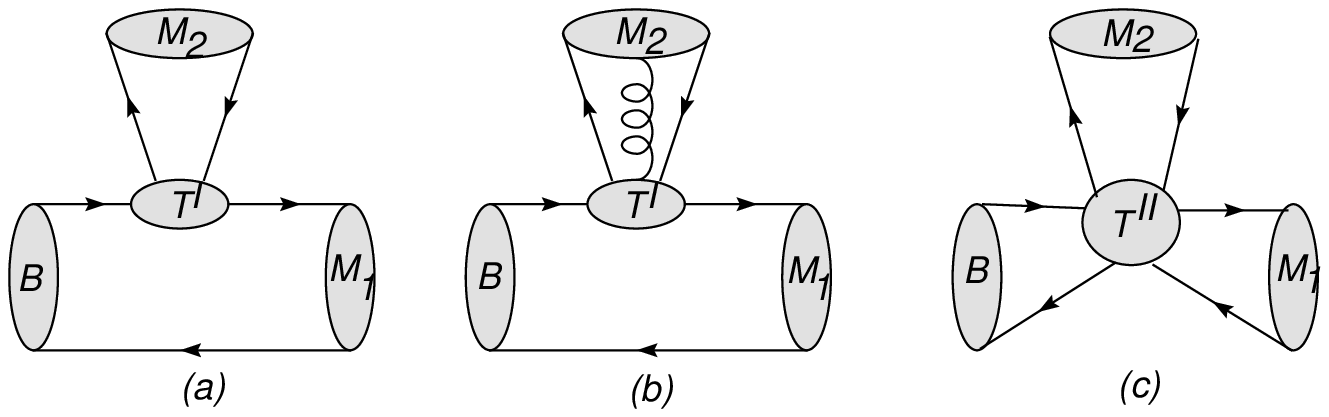}
\caption{\label{fig:fig1}The parton topologies correspond to the parton amplitudes of four, 
five and six parton interactions, respectively.}
\end{figure*}

To begin with, we first express the matrix element of an operator $O_i$ of the weak effective Hamiltonian $H_{eff}$ 
of the standard model for the hadronic decays $\bar{B}\to M_1 M_2$ 
in terms of parton model amplitudes
\bee\label{eq1}
&&\langle  M_1 M_2)|O_i|\bar{B}\rangle \nn
&=&\sum_{j=+,0} F^{B\to M_1}_j(m_{M_2}^2)\int\frac{d^4 l}{(2\pi)^4} 
\text{Tr}[T_{ij}^{I}(l)\Phi_{M_2}(l)]+(M_{1}\leftrightarrow M_{2})\nn
& & + \sum_{j=+,0} F^{B\to M_1}_j(m_{M_2}^2)\int\frac{d^4 l_1}{(2\pi)^4}\int\frac{d^4 l_2}{(2\pi)^4} 
\text{Tr}[T_{ij,\mu}^{I}(l_1,l_2)\Phi_{M_2}^{\mu}(l_1,l_2)]+(M_{1}\leftrightarrow M_{2})\nn
& & + \int\frac{d^4 l_B}{(2\pi)^4}\frac{d^4 l_{M_1}}{(2\pi)^4}\frac{d^4 l_{M_2}}{(2\pi)^4}
\text{Tr}[T_{i}^{II}(l_N,l_{M_1},l_{M_2})\Phi_{B}(l_B)\Phi_{M_1}(l_{M_1})\Phi_{M_2}(l_{M_2})]
 \;, 
\eee 
where $F_{j}^{B\to M_1}$ denote the form factors for $\bar{B}\to M_1 l\bar{\nu}$ transition.
The form factor $F_{j}$ are defined as
\bee
\langle M_1(p)|\bar{q}\gamma_{\mu}b|\bar{B}(P_b)\rangle=F_{+}(q^2)(p+P_b)_{\mu}+\frac{F_{+}(q^2)-F_{0}(q^2)}{q^2}q_{\mu}
\eee
where $q=P_b-p$ and $F_{+}(q^2)=F_{0}(q^2)$ under the limit ${q^2\to 0}$.
The parton amplitudes $T_{ij}^{I}(l)$, $T^{I}_{ij,\mu}(l_1, l_2)$ and $T^{II}_{i}(l_B, l_{M_1}, l_{M_2})$
are defined to describe the hard scattering center involving four parton, five parton, and six parton interactions
corresponding to those diagrams depicted in Fig.~\ref{fig:fig1}(a)-(c), respectively.
Note that there are also other types of parton amplitudes involving five or six parton interactions 
not being presented in Fig.~\ref{fig:fig1},
which can be attributed to either the physical form factors, or to be of higher twist than three. 
We have neglected these contributions in Eq.~(\ref{eq1}).
The $\text{Tr}$ symbol denotes the trace operation applied on the color and spin indices.
For convenience, we employ the light-cone coordinate system such that  
$P_{B}^{\mu}=(p^{\mu}+q^{\mu})$ with two light-like vectors $q^{\mu}=(q^+,q^-,q^{i}_{\perp})=(Q,0,0)$
and $p^\mu=(p^+,p^-,p^i_{\perp})=(0,Q,0)$ with $Q=m_B/\sqrt{2}$,
which are defined as the momenta carried by the final state $M_2$ and $M_1$ mesons, respectively.
The $M_1$ meson is defined to receive the spectator quark of the bottom meson.
The $M_2$ meson is defined as the emitted meson produced from the hard scattering center.
The hadron amplitudes $\Phi_{M_2}$ and $\Phi_{M_2}^{\mu}(l_1,l_2)$ are defined as
\bee
\Phi_{M_2}(l)&=&\int d^4 y  e^{il\cdot y } \langle M_2|\bar{q}(y)q(0)|0\rangle\;,\\
\Phi_{M_2}^{\mu}(l_1,l_2)&=&\int d^4 y  \int d^4 z e^{il_{1}\cdot y}e^{i (l_{2}-l_1)\cdot z}
 \langle M_2|\bar{q}(y)(-gA^{\mu}(z)) q(0)|0\rangle\;.
\eee
In our language, Eq.~(\ref{eq1}) contain leading, sub-leading and higher twist contributions. 
The Eq.~(\ref{eq1}) becomes meaningful only if the leading twist contributions can be separated 
from the sub-leading twist contributions.
For this purpose, we employ the Qiu's Feynman-diagram collinear expansion approach \cite{Qiu:1988dn}
to expand each Feynman diagram in a twist by twist manner.
As the loop corrections are considered, the twist expansion then interplays with the expansion in $\alpha_s$.
To be specific, we choose the following expansion strategy.
\begin{enumerate}
\item {All possible Feynman diagrams ordered in $\alpha_s$ are first drawn.}
\item {According to the collinear expansion (developed below), 
each Feynman diagram of order $O(\alpha_s^n)$ with $n\ge 0$  is expanded into a series ordered by increasing twist.}
\item {The contributions of the same twist order from the expansion series 
of each Feynman diagram with the same $\alpha_s$ order are added up together.}
\item{The factorization properties of the final expression with a specific twist and a specific $\alpha_s$ order are 
analyzed.}
\end{enumerate}  
The last one is important for us to derive a meaningful perturbation theory beyond the leading twist. 

For latter uses, we define the soft, collinear and hard loop parton momenta.
We let the soft momentum scale as $(l^+,l^-,l_\perp)\sim (\lambda,\lambda,\lambda)$,
the collinear momentum scale as $(l^+,l^-,l_\perp)\sim (Q,\lambda^2/Q,\lambda)$,
and the hard momentum scale as $(l^+,l^-,l_\perp)\sim (Q,Q,Q)$.
The scale variables are defined as $Q\sim m_b$ and $\lambda\sim\Lambda_{QCD}$.
For a collinear loop parton, 
it is convenient to parametrize its momentum $l^{\mu}$ into its components proportional to the meson momentum $q^{\mu}$,
the light-cone vector $n^{\nu}$, and the transversal directions
\bee\label{eq3}
l^{\mu} &=& n\cdot l q^{\mu} + \frac{l^2+ l_{\perp}^2}{2 n\cdot l}n^{\mu}+l^{\mu}_{\perp}\;,
\eee
where the vector $n^{\mu}$ satisfies $n\cdot q=1$, $n\cdot l_{\perp}=0$, and $n^2=0$. 
For convenience, we further define the collinear component, $\hat{l}^{\mu}$, 
the on-shell component, $l_{L}^{\mu}$, and the off-shell component, $l_{S}^{\mu}$ of the momentum $l^{\mu}$ as
\bee\label{momentum-expand}
\hat{l}^{\mu}&=& n\cdot l q^{\mu}\;,\nn
l^{\mu}_{L}&=& \hat{l}^{\mu}+\frac{l_{\perp}^2}{2n\cdot l}n^{\mu}+l_{\perp}^{\mu}\;,\nn
l_{S}^{\mu}&=&\frac{l^2}{2n\cdot l}n^{\mu}\;.
\eee
In the above expansion of the parton momentum into different parts, 
we have assumed $m_{M_{i}}=0$, $i=1,2$, and $q^2=0$. 
The loop partons except of the bottom quark are assumed massless for simplicity. 
The contributions from non-vanishing light quark masses are taken as corrections.
Because the light quark mass contributions are relatively negligible as compared to the bottom quark mass,
we also neglect the light quark mass effects in the following calculations. 

According to the parametrization in Eq.~(\ref{eq3}),
a parton propagator can be separated into its long distance part and short distance part (the special propagator).
If we write the loop parton propagator as \cite{Qiu:1988dn}
\bee\label{lp1}
F(y,z)&=&\int\frac{d^4 l}{(2\pi)^4}e^{il\cdot (y-z)}[F_L(l)+F_S(l)]\nn
      &=& F_L(y,z)+ F_S(y,z)\;,
\eee
where
\bee
F_L(l)=\frac{i\s{l}_{L}}{l^2}\;,\;\;
F_S(l)=\frac{i \s{n}}{2n\cdot l}\;.
\eee
The  $F_{L}(l)$ propagator corresponds to the long distance part of the propagator,
since $F_L(y,z)\propto \theta(y-z)$. 
The $F_{S}(l)$ propagator represents the short distance part because $F_S(y,z)\propto \delta(y-z)$.
We now describe one important property of the long distance propagator $F_{L}(l)$.
In a parton amplitude, the $F_{L}(l)$ may contact with a $\s{q}n^{\mu}$ component of a vertex $\gamma^{\mu}$. 
Whenever this happens,
the $\s{q}n^{\mu}$ vertex will extract one short distance propagator $F_{S}(l)$ 
and one interaction vertex $i\gamma_{\nu}$ from the relevant hadron amplitude \cite{Qiu:1988dn}
\bee\label{long-prop-1}
\frac{i\s{l}_{L}}{l^2}\s{q}=\frac{i\s{l}_L}{l^2}(i\gamma_{\nu})\frac{i\s{n}}{2n\cdot l}\s{q}(l-\hat{l})^{\nu}\;.
\eee
The momentum factor $(l-\hat{l})^{\nu}$ is then absorbed by the hadron amplitude due to the Ward identity 
\cite{Qiu:1988dn}.
We now explain how the identity Eq.~(\ref{long-prop-1}) can be obtained by a simple manipulation.
We first insert an identity $1=(\s{l}^2)/l^2$ into the left hand side of Eq.~(\ref{long-prop-1})
and expresse each $\s{l}$ into $\s{l}_L+\s{l}_S$ to obtain 
\begin{eqnarray}\label{long-prop-2}
\frac{i\s{l}_{L}}{l^2}\frac{\s{l}\s{l}}{l^2}\s{q}
=\frac{i\s{l}_{L}}{l^2}\frac{(\s{l}_L +\s{l}_S)(\s{l}_L +\s{l}_S)}{l^2}\s{q}\;.
\end{eqnarray}
Since $(\s{l}_L)^2=0=(\s{l}_S)^2$,
the above equation then becomes
\begin{eqnarray}\label{long-prop-3}
\frac{i\s{l}_{L}}{l^2}\frac{\s{l}\s{l}}{l^2}\s{q}
=\frac{i\s{l}_{L}}{l^2}\frac{(\s{l}_L \s{l}_S +\s{l}_S \s{l}_L)}{l^2}\s{q}\;,
\end{eqnarray} 
where the first term $\s{l}_L \s{l}_S$ in the right hand side  
leads to a vanishing result as it contacts with the term ${i\s{l}_{L}}/{l^2}$ term.
The only contribution can only come from the second term $\s{l}_S \s{l}_L$ in the right hand side of 
Eq.~(\ref{long-prop-3}).
In addition, the $ \s{l}_L$ can be expanded in the terms proportional to $\s{q}$, $\s{n}$, $\s{l}_{\perp}$.
This gives 
\[
\s{l}_S \s{l}_L\s{q}=l^2\frac{\s{n}}{2n\cdot l}(n\cdot l \s{q} + \frac{l_{\perp}^2\s{n}}{2n\cdot l}
+\s{l}_{\perp})\s{q}\;.
\]
Due to $\s{q}^2=\s{n}^2=0$, it further reduces to
\[
l^2\frac{\s{n}}{2n\cdot l}(\s{l}_{\perp})\s{q}\;.
\]
By substituting the above back into Eq.~(\ref{long-prop-2}), 
Eq.~(\ref{long-prop-1}) is then obtained by noting that 
\[
\s{l}_{L}(i\gamma_{\alpha})(i\s{n})\s{q}(l-\hat{l})^{\alpha}=\s{l}_{L}\s{n}\s{l}_{\perp}\s{q}\;.
\]
Using Eq.~(\ref{long-prop-1}), one can systematically include the effects from the non-collinearity 
and the off-shellness of the collinear partons.
This property of the long distance part of the parton propagator plays an important role in our following analysis,
and its meaning will become more clear after we have investigated real cases lately. 

According to the parton model,
the hadron amplitudes are defined as the probability for finding the on-shell partons inside the hadron.
The parton amplitudes are then required to contain only the on-shell components of 
the external parton momenta.
However, according to Eq.~(\ref{momentum-expand}),
either the on-shell momentum $l_L$ or the collinear momentum $\hat{l}$ can be assigned for an on-shell parton.
Therefore, there arise two factorization schemes, 
the collinear factorization \cite{Lepage:1979zb,Lepage:1980fj,Efremov:1979qk,Chernyak:1977fk,Chernyak:1980dj,Chernyak:1983ej}(QCD factorization)
and the $k_{T}$ factorization \cite{Catani:1990xk,Catani:1990eg,Collins:1991ty,Levin:1991ry}
(PQCD factorization 
\cite{Li:1992nu,Nagashima:2002ia,Li:1994cka,Li:1995jr,Li:1994iu,Li:1996ds,Chang:1996dw,Yeh:1997rq,Keum:2000ph,Keum:2000wi,Keum:2000ms}).
In the $k_{T}$ factorization scheme, an on-shell parton carries a momentum $l_L$. 
On the other hand, in the collinear factorization scheme, 
an on-shell parton carries a momentum $\hat{l}$.
In this work, we follow the QCD factorization to use the collinear factorization scheme as our basics.
Our proposed collinear expansion method is composed of following steps:
\begin{enumerate}
\item Use scale analysis for the parton amplitudes according to the scales of parton momenta to find out the 
leading regions of the parton momentum configuration.
\item The parton amplitudes are expanded into a Taylor series with respect to the leading regions of parton momenta.
\item The expanded parton amplitudes are substituted back into the contraction with the hadron amplitudes to extract
relevant contributions up to specific twist order.
\item The factorization of parton momentum integrals is accomplished by means of 
integral transformations (See, for example,  Eq.~(\ref{integral-transf})). 
\item The color structure of the parton amplitude is extracted to be attributed to the hadron amplitudes to complete
the color factorization.
\item The factorization of spin indices is completed by means of Fierz transformation.
\item The property of the long distance parton propagator is used to extract higher twist  contributions.
\end{enumerate}

\begin{figure*}
\includegraphics{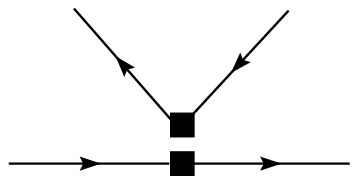}
\caption{\label{fig:fig2}The Feynman diagram describes the tree level four parton amplitude, $T^{I(0)}_{ij}$. 
The square symbol represents the vertex of weak interactions.}
\end{figure*}

We are now ready to discuss the collinear expansion.
First, we order the parton amplitudes in $\alpha_s$
\bee
&&T^{I}_{ij}(l)=T^{I(0)}_{ij}+T^{I(1)}_{ij}(l)+O(\alpha_s^2)\;,\\
&&T^{I}_{ij,\mu}(l_1,l_2)=T^{I(0)}_{ij,\mu}(l_1,l_2)+T^{I(1)}_{ij,\mu}(l_1,l_2)+O(\alpha_s^2)\;,\\
&&T^{II}_{i}(l_B, l_{M_1}, l_{M_2})=T^{II(1)}_{i}(l_B, l_{M_1}, l_{M_2})+O(\alpha_s^2)\;,
\eee
where the superscription $(0)$ and $(1)$ are used to denote the relevant parton amplitude of zeroth and first order in
$\alpha_s$, respectively. 
The parton amplitude $T^{I(0)}_{ij}$ is just the tree vertex $\bar{\Gamma}_i\delta_{ij}$ 
in the diagram as depicted in Fig.~\ref{fig:fig2}.
There are vertex and penguin diagrams for $T^{I(1)}_{ij}$ amplitudes as depicted in Fig.~\ref{fig:fig4}.
The parton amplitude $T^{I(0)}_{ij,\mu}$ describes the tree diagrams as depicted in Fig.~\ref{fig:fig3},
in which two quark partons and one gluon parton from the meson $M_2$ are interacting with a 
local four fermion operator $O_i=(\bar{q}_1 \Gamma_i b)(\bar{q}_2\bar{\Gamma}_i q_3)$ with 
$\Gamma_i(\bar{\Gamma}_i)$ the Dirac gamma matrix for the operator.
The parton amplitude $T^{II}_{i}$ starts from $O(\alpha_s)$ diagrams as depicted in Fig.~\ref{fig:fig5}
and is denoted as $T^{II(1)}$.

We first expand the parton amplitudes $T^{I(0)}_{ij,\mu}$ with respect to the
collinear components of their relevant parton momenta as
\bee
&&T^{I(0)}_{ij,\mu}(l_1,l_2)=T^{I(0)}_{ij,\mu}(\hat{l}_1,\hat{l}_2)
+\sum_{k=1}^2\left.\frac{\partial T^{I(0)}_{ij,\mu}}{\partial l_k^{\nu}}\right|_{l_k=\hat{l}_k}(l_k -\hat{l}_k)^{\nu}+\cdots\;,
\eee 
where
\bee
&& T_{ij,\mu}^{I(0)}(\hat{l}_{1},\hat{l}_{2})=((i\gamma_\mu)\frac{i\s{\hat{l}}_2}{\hat{l}^2_2}\bar{\Gamma}_i 
+\bar{\Gamma}_{i}\frac{-i \s{\bar{\hat{l}}}_1}{\bar{\hat{l}}_{1}^2}(-i\gamma_\mu))\delta_{ij}\;,\\
&& \frac{\partial T^{I(0)}_{ij,\mu}}{\partial l_k^{\nu}}(\hat{l}_{1},\hat{l}_{2})
= ((i\gamma_\mu)\frac{i\s{\hat{l}}_2}{\hat{l}^2_2}(i\gamma_\nu)\frac{i\s{\hat{l}}_2}{\hat{l}^2_2}\bar{\Gamma}_i 
\delta_{k2}
-\bar{\Gamma}_{i}\frac{-i \s{\bar{\hat{l}}}_1}{\bar{\hat{l}}_{1}^2}(-i\gamma_\nu )
\frac{-i \s{\bar{\hat{l}}}_1}{\bar{\hat{l}}_{1}^2}(-i\gamma_\mu)\delta_{k1})\delta_{ij}\;.
\eee
The expansion series are then substituted back into the convolution integrals 
with the hadron amplitudes for further analysis.
The reason why one can expand the parton amplitudes with respect 
to the relevant collinear momenta will be explained in detail below.
\subsection{Expansion with $T_{ij}^{I(0)}$}
The expression for the contributions associated with $T_{ij}^{I(0)}$ is written as
\bee
\sum_{j=+,0} F^{B\to M_1}_{j}(m_{M_2}^2)\int\frac{d^4 l}{(2\pi)^4} 
\text{Tr}[T_{ij}^{I(0)}\Phi_{M_2}(l)]\;,
\eee
where the loop momentum $l$ is carried by the loop parton in Fig.~\ref{fig:fig1}(a).
Since the parton amplitude $T_{ij}^{I(0)}$ is independent of $l$, 
we propose to use the following integral identity to transform the expression into a form
consistent with the parton model picture
\bee\label{integral-transf-1}
&&\int_0^1 dx \delta(x-n\cdot l)\nn
&&=\int_0^1 dx 
\int_{0}^{\infty}\frac{d\lambda}{2\pi}e^{i\lambda(x- n\cdot l)}\nn
&&=1\;.
\eee
The transformed result appears as
\bee
\int_0^1 dx \text{Tr}[T_{ij}^{I(0)}\Phi_{M_2}(x)]
\eee
where
\bee
\Phi_{M_2}(x)=\int_{0}^{\infty}\frac{d\lambda}{2\pi} e^{i\lambda x}
\langle M_2|\bar{q}(\lambda n)q(0)|0\rangle\;.
\eee
The following integral transformation has been used in the above to simplify the expression
\bee\label{integral-transf}
&&\int\frac{d^4 l}{(2\pi)^4}\int d^4 y e^{il\cdot (y-\lambda n)} G(y,0)\nn
&&=\int d^4 y \delta^{(4)}(y-\lambda n)G(y,0)\nn
&&= G(\lambda n,0)\;,
\eee
where $G(y,0)$ denotes any function of the coordinates.
Two comments for the above integral transformations Eqs.~(\ref{integral-transf-1}) 
and (\ref{integral-transf}) are necessary.
First, the momentum fraction $x$ for the parton of the meson $M_2$ is introduced.
Second, the quark field $\bar{q}(\lambda n)$ is ordered in light-cone direction $n$.
This implies that the hadron amplitude $\Phi_{M_2}(x)$ is defined on the light-cone $n^2=0$ 
where $n^{\mu}$ is a null light-cone vector. 
By using the above integral transformations, 
the parton amplitude and the hadron amplitude are only related by the momentum fraction $x$.
Because the parton amplitude $T_{ij}^{I(0)}$ is equal to $\bar{\Gamma}_i\delta_{ij}$,
the integral over $x$ is then associated with $\Phi_{M_2}(x)$.

The factorization of spin indices depends on the structure of $\bar{\Gamma}_i$.
For $(V-A)(V\pm A)$ operators, $\bar{\Gamma}_i=(V\pm A)$ and can be expanded into 
$\s{q}(1\pm\gamma_5)$, $\s{n}(1\pm\gamma_5)$ and $\gamma_{\perp}\gamma_5$. 
If $M_2$ is a pseudo-scalar meson,
only the axial vector part can contribute.
However, only $\s{n}\gamma_5$ leads to leading twist contributions.
The $\s{q}\gamma_5$ will result in twist-4 contributions and $\gamma_{\perp}\gamma_5$ will not contribute.
For other types of meson, similar considerations can be made. 
We now explain how the $\s{q}\gamma_5$ part can contribute.
The long distance part of the parton propagator can interact with vertex $\s{q}\gamma_5$ to have
\bee
\frac{i\s{l}_{L}}{l^2}\s{q}=\frac{i\s{l}_L}{l^2}(i\gamma_{\alpha})\frac{i\s{n}}{2n\cdot l}\s{q}(l-\hat{l})^{\alpha}\;.
\eee
It is also applicable for the other parton propagator of the anti-quark line. 
The short distance part of the parton propagator $i\s{n}/(2n\cdot l)$ and the vertex $i\gamma_\alpha$ are 
absorbed into the parton amplitude.
This results in
\bee\label{twist4-1}
\int_0^1 dx \text{Tr}[T_{ij,\alpha\beta}(x,x,x)w^{\alpha}_{\alpha^\prime}w^{\beta}_{\beta^\prime}
\Phi_{M_2,\partial}^{\alpha^\prime\beta^\prime}(x,x,x)]
\eee
where $w^{\alpha}_{\alpha^\prime}=g^{\alpha}_{\alpha^\prime}-q^{\alpha}n_{\alpha^\prime}$, 
\bee
&&T_{{ij},\alpha\beta}(x,x,x)\equiv 
(i\gamma_\alpha)\frac{i\s{n}}{2n\cdot l}T^{I(0)}_{ij}\frac{-i\s{n}}{2 n\cdot \bar{l}}(-i\gamma_\beta )\;,\nn
&&\Phi_{M_2,\partial}^{\alpha^\prime\beta^\prime}(x,x,x)
=\int_{0}^{\infty}\frac{d\lambda}{2\pi} e^{i\lambda x}
\langle M_2|\bar{q}(\lambda n)i\partial^{\alpha^\prime}(\lambda n)i\partial^{\beta^\prime}(\lambda n)q(0)|0\rangle\;.
\eee
There are corresponding contributions from the two gluon insertion diagrams depicted in Fig.~\ref{fig:fig6},
whose expression is written as
\bee
\int \frac{d^4 l_1}{(2\pi)^4}
\frac{d^4 l_2}{(2\pi)^4}
\frac{d^4 l_3}{(2\pi)^4}
\text{Tr}[T^{I(0)}_{ij,\alpha\beta}(l_1,l_2,l_3)w^{\alpha}_{\alpha^\prime}w^{\beta}_{\beta^\prime}
\Phi_{M_2,A}^{\alpha\beta}(l_1,l_2,l_3)]
\eee
where we have employed the light-cone gauge $n\cdot A=0$ for the gluon fields, and the parton amplitude and
hadron amplitude are expressed as
\bee
&&T_{ij,\alpha\beta}^{I(0)}(l_1,l_2,l_3)=
(i\gamma_\alpha)\frac{i\s{n}}{2n\cdot l_2}T^{I(0)}\frac{-i\s{n}}{2 n\cdot \bar{l}_2}(-i\gamma_\beta )\nn
&&\Phi_{M_2,A}^{\alpha\beta}(l_1,l_2,l_3)
=\int d^4 z \int d^4 y \int d^4 w
 e^{il_1\cdot y}e^{i(l_2-l_1)\cdot z} e^{i (l_3-l_2)\cdot w} \nn
&&\times\langle M_2|\bar{q}(y)(-gA^{\alpha}(z))(-gA^{\beta}(w))q(0)|0\rangle
\eee
Since $T_{ij,\alpha\beta}^{I(0)}(l_1, l_2,l_3)$ can be replaced by $T_{ij,\alpha\beta}^{I(0)}(x_1, x_2,x_3)$
straightforwardly,
the momentum integrations over $l_1, l_2,l_3$ can be transformed into the integrations over $x_1, x_2,x_3$.
We then obtain
\bee\label{twist4-2}
\int dx_1 dx_2 dx_3
\text{Tr}[T^{I(0)}_{ij,\alpha\beta}(x_1,x_2,x_3)\Phi_{M_2,A}^{\alpha\beta}(x_1,x_2,x_3)]\;.
\eee
The combination of Eq.~(\ref{twist4-1}) and Eq.~(\ref{twist4-2}) gives
\bee
\int dx_1 dx_2 dx_3
\text{Tr}[T^{I(0)}_{ij,\alpha\beta}(x_1,x_2,x_3)w^{\alpha}_{\alpha^\prime}w^{\beta}_{\beta^\prime}
\Phi_{M_2,D}^{\alpha^\prime\beta^\prime}(x_1,x_2,x_3)]
\eee
where
\bee
\Phi_{M_2,D}^{\alpha\beta}(x_1,x_2,x_3)&=&
\int \frac{d\lambda}{2\pi} \int \frac{d\eta}{2\pi} \int\frac{d\omega}{2\pi}
e^{i \lambda x_1}e^{i \eta(x_2-x_1)} e^{i \omega(x_3-x_2)} \nn
&&\times\langle M_2|\bar{q}(\lambda n)(iD^{\alpha}(\eta n) )(iD^{\beta}(\omega n ))
q(0)|0\rangle
\eee
with $iD^{\alpha}=i\partial^\alpha -gA^{\alpha}$ being the covariant derivative.
Since $\s{n}$ is of $O(Q^{-1})$, $T^{I(0)}_{ij,\alpha\beta}$ is of $O(Q^{-2})$ as the scale of 
$T^{I(0)}_{ij}$ being of $O(1)$.
The relevant contributions are of higher than twist-4.
The above example is to show that, 
using the collinear expansion, 
one can calculate the tree level higher twist corrections from the dynamical partons in a systematic way.
Because we only intend to calculate the twist-3 corrections,
we will not further explore the contributions of twist order higher than three.
For $-2(S-P)(S+P)$ operators, $\bar{\Gamma}_i=\gamma_5$.
Up to twist-3, the expression appears as
\bee
\int_0^1 dx \text{Tr}[T_{ij}^{I(0)}\Phi_{M_2}(x)]
&=&
-\frac{1}{4}\int_0^1 dx \text{Tr}[T_{ij}^{I(0)}\s{q}\gamma_5 ]\text{Tr}[\s{n}\gamma_5\Phi_{M_2}(x)]\nn
&&+\frac{1}{4}\int_0^1 dx \text{Tr}[T_{ij}^{I(0)}\gamma_5]\text{Tr}[\gamma_5\Phi_{M_2}(x)]\;.
\eee
By identifying $\text{Tr}[\s{n}\gamma_5\Phi_{M_2}(x)]$ and
$\text{Tr}[\gamma_5\Phi_{M_2}(x)]$ as the twist-2 and twist-3 two parton LCDAs of the $M_2$ meson
\bee
\text{Tr}[\s{n}\gamma_5\Phi_{M_2}(x)]&=& -if_{M_2}\phi_{M_2}^{tw2}(x)\;,\\
\text{Tr}[\gamma_5\Phi_{M_2}(x)]&=& -if_{M_2}\mu_{\chi}\phi_{M_2,P}^{tw3}(x)\;,
\eee 
where $\mu_{\chi}^{M_2}=m_{M_2}^2/(\bar{m}_q+ \bar{m}_{\bar{q}})$ with $\bar{m}_{q}$ and $\bar{m}_{\bar{q}}$
the current quark masses and $m_{M_2}$ the meson mass,
we recover the naive factorization result up to twist-3 order.

%
\subsection{Expansion with $T_{ij,\mu}^{I(0)}$}
We show the light-cone gauge $n\cdot A=0$ and the covariant gauge $\partial\cdot A=0$ 
for the expansion with $T_{ij,\mu}^{I(0)}$.
Because the analysis is tedious, we outline the procedure, here, 
and leave the details for next section.
The first step is to take a power counting for the parton amplitude $T_{ij,\mu}^{I(0)}$. 
There are three interesting regions. 
%
The first region is composed of either two soft loop parton momenta, or one soft loop parton momentum and one collinear loop parton momentum.
The $T_{ij,\mu}^{I(0)}$ in the first region is counted as $\lambda^{-1}$.
The second region is composed of two collinear loop parton momenta.
The $T_{ij,\mu}^{I(0)}$ in the second region is counted as $Q\lambda^{-2}$.
The third region is composed of either one collinear loop parton momentum and one hard loop parton momentum, 
or two hard loop parton momenta. 
In the third region, the $T_{ij,\mu}^{I(0)}$ is counted as $Q^{-1}$.
We conclude that the region composed of two collinear loop parton momenta is dominant.
 
Let's, first, consider the light-cone gauge $n\cdot A=0$.
The expansion series of $T_{ij,\mu}^{I(0)}(l_1,l_2)$ with respect to $\hat{l}_i$, $i=1,2$, are written as
\bee\label{expansion-1}
T_{ij,\mu}^{I(0)}(l_1,l_2)=T_{ij,\mu}^{I(0)}(\hat{l}_1,\hat{l}_2)
+\sum_{k=1,2}T_{ijk,\mu\nu}^{I(0)}(\hat{l}_1,\hat{l}_k,\hat{l}_2)(l_k -\hat{l}_k)^{\nu }+\cdots\;,
\eee 
where $T_{ijk,\mu\nu}^{I(0)}(\hat{l}_1,\hat{l}_k,\hat{l}_2)$ are defined by assuming the low energy theories
\bee
T_{ijk,\mu\nu}^{I(0)}(\hat{l}_1,\hat{l}_k,\hat{l}_2)
=\left.\frac{\partial T_{ij,\mu}^{I(0)}}{\partial l_k^{\nu}}\right|_{l_1=\hat{l}_1, l_2=\hat{l}_2}\;.
\eee 
The expansion series are then substituted back into the convolution integrals.
Since the gauge condition $n\cdot A=0$ with a light-cone vector $n^{\mu}$ satisfying $n\cdot q=1$,
$n^2=0$, and $n\cdot l_{\perp}=0$,
the first term $T_{ij,\mu}^{I(0)}(\hat{l}_1,\hat{l}_2)$ leads to the result
\bee
\int\frac{d^4 l_1}{(2\pi)^4}\int\frac{d^4 l_2}{(2\pi)^4}
\text{Tr}[T_{ij,\mu}^{I(0)}(\hat{l}_1,\hat{l}_2)w^{\mu}_{\alpha}\Phi_{M_2}^{\alpha}(l_1,l_2)]
\eee
where we have introduced $w^{\mu}_{\alpha}=g^{\mu}_{\alpha}-q^{\mu}n_{\alpha}$.
Similarly, we employ the following integral identities to simplify the momentum integrations
\bee
\int_0^1 dx_1 \int_0^1 dx_2 \int_0^{\infty} \frac{d\lambda}{2\pi}\int_0^{\infty}\frac{d\eta}{2\pi}
e^{i\lambda (x_1 -n\cdot l_1)} e^{i\eta (x_2 - n\cdot l_2)}=1
\eee
The expression appears as
\bee
\int_0^1 dx_1 \int_0^1 dx_2\text{Tr}[T_{ij,\mu}^{I(0)}(x_1,x_2)w^{\mu}_{\alpha}\Phi_{M_2}^{\alpha}(x_1,x_2)]
\eee
in which
\bee
T_{ij,\mu}^{I(0)}(x_1,x_2)&=& \left. T_{ij,\mu}^{I(0)}(\hat{l}_1,\hat{l}_2)\right|_{\hat{l}_1=x_1,\hat{l}_2=x_2}\;,\\
\Phi_{M_2}^{\alpha}(x_1,x_2)&=&
\int_0^{\infty} \frac{d\lambda}{2\pi}\int_0^{\infty}\frac{d\eta}{2\pi}
e^{i\lambda x_1} e^{i \eta ( x_2-x_1)} \nn
&& \times\langle M_2|\bar{q}(\lambda n)(-gA^{\mu}(\eta n)) q(0)|0\rangle\;.
\eee
In the above equations, we have used the following transformation for any function $G(y,z,0)$ 
of coordinates $y^{\mu}$ and $z^{\mu}$
\bee
&&\int\frac{d^4 l_1}{(2\pi)^4}\int\frac{d^4 l_2}{(2\pi)^4}\int d^4 y \int d^4 z
 e^{i l_1\cdot(y-z- \lambda n)}e^{i l_2\cdot(z- \eta n) } G(y,z,0)\nn
&&=G(\lambda n, \eta n,0)\;. 
\eee

For tree diagrams, only color singlet operators can contribute at twist-3.
The remaining task is to finish the factorization of spin indices.
For $(V-A)(V\pm A)$ operators , 
the related contributions are of twist-4,
which are beyond our accuracy.
For $-2(S-P)(S+P)$ operators, the related contributions are of twist-3.
To obtain the collinear limit $\hat{l}_i \rightarrow x_i$, $i=1,2$, for $T_{ij,\mu}^{I(0)}(x_1,x_2)$ 
, we have used the following substitution for the parton propagators  
\bee
\frac{i\hat{\s{l}}_1}{\hat{l}_1^2}\rightarrow\frac{i\s{n}}{2n\cdot l_1}\;.
\eee
This is because the vertex $i\gamma_{\mu}$ in $T_{ij,\mu}^{I(0)}(x_1,x_2)$ is transversal and
only the off-shell part of the parton propagators can contribute. 
See further explanations in the next section.
The terms associated with $T^{I(0)}_{ijk,\mu\nu}$ are of twist-4 and higher.
They are neglected accordingly. 
The factorization of spin indices results in
\bee\label{final-1}
\frac{1}{8}\int_0^1 dx_1 \int_0^{\bar{x}_1} dx_2 
\text{Tr}[T^{I(0)}_{ij,\mu}\sigma_{\alpha\beta}\gamma_5]w^{\mu}_{\mu^{\prime}}
\text{Tr}[\sigma^{\alpha\beta}\gamma_5\Phi_{M_2}^{\mu^{\prime}}(x_1,x_2)]\;.
\eee
The explicit expression for Eq.~(\ref{final-1}) is left to the next section. 

We now consider the expansion with covariant gauge $\partial\cdot A =0$.
Since the factorizations of the momentum integrals and color indices are independent of gauge condition,
we can go through to consider the factorization of spin indices.
The first term in the expansion of $T_{ij,\mu}^{I(0)}(x_1,x_2)$, under covariant gauge, 
is related to the gauge invariant phase factor of the related two parton amplitudes.
The gluon fields $A^{\mu}$ in $\Phi_{M_2}^{\mu}$ can be expanded as 
$A^{\mu}=n\cdot A q^{\mu}+q\cdot A n^{\mu}+ d^{\mu}_{\alpha}A^{\alpha}$.
The contraction $\text{Tr}[T_{ij,\mu}^{I(0)}(x_1, x_2)q^{\mu}n\cdot\Phi_{M_2}(x_1,x_2)]$ leads to
\bee
\int dx_1 \int dx_2 
\text{Tr}[ \bar{\Gamma}_i\frac{2}{n\cdot k}n\cdot\Phi_{M_2}(x_1,x_2)]\;,
\eee 
where $k=l_2-l_1$ being the gluon momentum
and
\bee
n\cdot\Phi_{M_2}(x_1,x_2)=\int_0^{\infty} \frac{d\lambda}{2\pi}\int_0^{\infty}\frac{d\eta}{2\pi}
e^{i\lambda x_1} e^{i \eta (x_2-x_1)}
\langle M_2|\bar{q}(\lambda n)(-g n\cdot A (\eta n)) q(0)|0\rangle\;.
\eee
The terms with $q\cdot A n^{\mu}$ vanish since the covariant gauge condition $\partial\cdot A=0$.
The terms with the contraction $\text{Tr}[T_{ij,\mu}^{I(0)}(x_1, x_2)d^{\mu}_{\alpha}\Phi_{M_2}^{\alpha}(x_1,x_2)]$
are of higher twist than twist-3 and will be neglected.
With the above considerations, the contraction $\text{Tr}[T_{ij,\mu}^{I(0)}(x_1, x_2)\Phi_{M_2}^{\mu}(x_1,x_2)]$
leads to contributions of twist-2 or higher than twist-3.

We next consider the contraction
$\text{Tr}[T_{ijk,\mu\nu}^{I(0)}(x_1, x_2)(l_k -\hat{l}_k)^{\nu}\Phi_{M_2}^{\mu}(x_1,x_2)]$,
which can be rewritten as
\bee
\text{Tr}[T_{ijk,\mu\nu}^{I(0)}(x_1,x_k, x_2)w^{\nu}_{\nu^\prime}\Phi_{M_2}^{\nu^\prime\mu}(x_1,x_k,x_2)]
\eee
with
\bee
\Phi_{M_2}^{\nu^\prime\mu}(x_1,x_k,x_2)\equiv \int\frac{d\lambda}{2\pi}\int\frac{d\eta}{2\pi}
e^{i\lambda x_1}e^{i\eta (x_2-x_1)}\langle M_2|\bar{q}_2(\lambda n)igG^{\nu^{\prime}\mu}(\eta n) q_3(0)|0\rangle\;.
\eee
Note that only transversal part $d^{\nu}_{\perp,\beta}(l_k -\hat{l}_k)^{\beta}$ of the $(l_k -\hat{l}_k)^{\nu}$
can contribute at twist-3.
For $(V-A)(V\pm A)$ operators, the contributions are of twist-4.
For $-2(S-P)(S+P)$ operators, the result appears as
\bee
&&\frac{1}{8}\int_0^1 dx_1 \int_0^{\bar{x}_1} dx_2\int dx_k 
\text{Tr}[T_{ijk,\mu\nu}^{I(0)}(x_1,x_k, x_2)d^{\nu}_{\perp,\nu^\prime}\sigma_{\alpha\beta}\gamma_5]
[\sigma^{\alpha\beta}\gamma_5\Phi_{M_2}^{\nu^\prime\mu}(x_1,x_k, x_2)]\nn
&&\times(\delta(x_k-x_1)+\delta(x_k-x_2))\;.
\eee
The reader may have noticed that the terms in the expansion series of $T_{ij,\mu}^{I(0)}(l_1,l_2)$ in 
Eq.~(\ref{expansion-1}) are of different twist order under the covariant or the light-cone gauge.
For example, the twist-3 contributions are from the the first term 
in the expansion series of $T_{ij,\mu}^{I(0)}(l_1,l_2)$ under the light-cone gauge.
On the other hand, under the covariant gauge, the twist-3 contributions are from the the second term 
in the expansion series of $T_{ij,\mu}^{I(0)}(l_1,l_2)$.
Since the parton amplitude and hadron amplitude under the collinear expansion are required to be gauge invariant,
 respectively,
this feature of the collinear expansion method can be used as a guiding principle for calculations.

\section{twist-3 corrections}
In this section, we make a more detail descriptions for the twist-3 contributions 
from the three parton Fock state $q\bar{q}g$ of the $M_2$ meson. 
The amplitude for the three parton $q\bar{q} g$ of $M_2$ interacting with the operator $O_6$ at the tree level for 
$\bar{B}\to M_1 M_2$ decays is written as
\bee\label{twist3-1}
&&\langle M_1|\bar{q}_1(0)(1-\gamma_5)b(0)|\bar{B}\rangle\nn
&&\times
\int d^4y \int d^4z \int\frac{d^4 l}{(2\pi)^4}\frac{d^4 k}{(2\pi)^4}e^{il\cdot z}e^{ik\cdot(y- z)}
\langle M_2|\bar{q}_2(z)[(-ig\s{A}(y))
\frac{i(\s{l}+\s{k})}{(l+k)^2+i\epsilon}(1+\gamma_5)\nn
&&+\frac{-i(\s{q}-\s{l}+\s{k})}{(q-l+k)^2+i\epsilon}(+ig\s{A}(y))(1+\gamma_5)]q_3(0)|0\rangle\; .
\eee
The $l$ and $k$ denote the momenta carried by the $q_2$ quark and $g$ gluon fields in Fig.~\ref{fig:fig3}(a) and (b). 
We first employ the light-cone gauge $n\cdot A(y)=0$. 
The gluonic fields $A^{\alpha}(y)$ represents $A^{\alpha,a}(y)T^a$ with the color matrix $T^{a}$ in the fundamental
representation $\sum T^a T^b =\delta^{ab}/2$. 
To relate to the previous introduced collinear expansion, 
we recast the convolution integration part of Eq.~(\ref{twist3-1}) into the form 
\bee\label{twist3-2}
\int\frac{d^4 l}{(2\pi)^4}\frac{d^4 k}{(2\pi)^4}\Tr[T^{I(0)}_{\mu}(k,l)w^{\mu}_{\mu^\prime}\Phi^{\mu^\prime}(k,l)]
\eee
where the parton amplitude $T^{I(0)}_{\mu}(k,l)$ is defined as
\bee
T^{I(0)}_{\mu}(k,l)=[(i\gamma_{\mu})\frac{i(\s{l}+\s{k})}{(l+k)^2+i\epsilon}
+\frac{-i(\s{q}-\s{l}+\s{k})}{(q-l+k)^2+i\epsilon}(-i\gamma_{\mu})](1+\gamma_5)
\eee
and the meson amplitude $\Phi^{\mu^\prime}(k,l)$
\bee
\Phi^{\mu^\prime}(k,l)=\int d^4y \int d^4z 
e^{il\cdot z}e^{ik\cdot(y- z)}
\langle M_2|\bar{q}_2 (z)(-gA^{\mu^\prime}(y))q_3(0)|0\rangle\ . 
\eee
The tensor $w^{\mu}_{\mu\prime}=g^{\mu}_{\mu\prime}-q^{\mu}n_{\mu\prime}$ 
has been introduced. 
Note that, for convenience, we have made a change of variables for the loop parton momenta $l=l_1^{\mu}$ 
and $k^{\mu}=(l_2 -l_1)^{\mu}$.
We assume that the emitted $M_2$ meson is highly energetic.
As shown in last section, the dominante configuration is composed of collinear $l_1$ and $l_2$. 
This allows us to expand the parton amplitude $T^{I(0)}_{ij,\alpha}(k,l)$ with respect to 
$\hat{l}=x q$ and $\hat{k}=(x^{\prime}-x)q$
\bee\label{twist3G-1}
T^{I(0)}_{ij,\mu}(k, l)=T^{I(0)}_{\mu}(\hat{k}, \hat{l})
+\left.\frac{\partial T^{I(0)}_{ij,\mu}(k,l)}{\partial l^\nu}\right|_{l=\hat{l}, k=\hat{k}}(l-\hat{l})^{\nu}
+\left.\frac{\partial T^{I(0)}_{ij,\mu}(k,l)}{\partial k^\nu}\right|_{l=\hat{l}, k=\hat{k}}(k-\hat{k})^{\nu}
+\cdots \ .
\eee
Substituting the first term back into the convolution integrations gives 
\bee
\int dx \int dx^\prime
\Tr[T^{I(0)}_{ij,\mu}(x,x^\prime)w^{\mu}_{\mu^\prime}
\Phi^{\mu^\prime}(x,x^\prime)]
\eee
where
\bee
\Phi^{\mu^\prime}(x^\prime,x)=
&&\int\frac{d^4 l}{(2\pi)^4}\delta(x-l\cdot n)
\int\frac{d^4 k}{(2\pi)^4}\delta(x^\prime-x-k\cdot n)
\int d^4y \int d^4z \nn
&&\times e^{il\cdot z}e^{ik\cdot y}
\langle M_2|\bar{q}_2(z)(-gA^{\mu^\prime}(y))q_3(0)|0\rangle\ .
\eee
In the above collinear limit step 
$T^{I(0)}_{ij,\mu}(l,k)\to T^{I(0)}_{ij,\mu}(x, x^{\prime})$,
there arises an infrared divergence as $x^\prime\to 0$,
which is from the denominators of virtual quark propagators 
\bee
\frac{ix^\prime \s{q}}{(x^\prime q)^2+i\epsilon}\ .
\eee
We regularize this divergence by the following method.
Since the full quark propagator with momentum $l^\prime = l+ k$ can be decomposed into its long distance part 
and short distance part as
\bee
\frac{i\s{l}}{(l+k)^2+i \epsilon}=\frac{i\s{l}^{\prime}_L}{(l^\prime)^2 +i\epsilon}+\frac{i\s{n}}{2n\cdot l^{\prime}}\;.
\eee
The long distance part gives vanishing result upto twist-3.
The short distance part is absorbed by the parton amplitude.
The divergence is then regularized by replacing the quark propagators with its corresponding
special propagators \cite{Qiu:1988dn, Boer:2001tx}
\bee
\frac{ix^\prime \s{q}}{(x^\prime q)^2+i\epsilon}\to
\frac{i\s{n}}{2x+i\epsilon}\frac{x^\prime-x}{x^\prime-x+i\epsilon}\;.
\eee  
The introduction of a special propagator for an on-shell fermion propagator is due to the fact that
the fermion propagators in Fig. 3(a) and 3(b) become on-shell and divergent after the collinear expansion.
The divergent part of these propagators leads to long distance contributions that should be included into
the twist-2 distribution amplitude for the $M_2$ meson.
However, there are also finite contact part of these propagators, which leads to contributions of one twist higher.
The more detailed explanation about the meaning of the special propagator refers to \cite{Qiu:1988dn, Boer:2001tx}.  

Under light-cone gauge $n\cdot A(y)=0$, 
it is convenient to transform the gluon fields $A^{\mu}(y)$ into its field strength $G^{\nu\mu}(y)$ by using
 following replacement
\bee
A^{\mu^\prime}(y)\to\frac{i n_{\nu}G^{\nu\mu^\prime}(y)}{(x^\prime-x)}\; ,
\eee
and
\bee
\Phi_{M_2}^{\mu^{\prime}}(x,x^\prime)\to \frac{in_{\nu}}{x^\prime-x}\Phi_{M_2}^{\nu\mu^{\prime}}(x,x^\prime)\; .
\eee
The factor $in_{\nu}/(x^\prime-x)$ is then absorbed by $T^{I(0)}_{\mu}(x,x^\prime)$ into the form
\bee
T^{I(0)}_{\mu\nu}(x,x^\prime)\equiv T^{I(0)}_{\mu}(x,x^\prime)\frac{in_{\nu}}{x^\prime-x}\;.
\eee

The factorization of the spin indices gives
\bee
\frac{1}{8}\int dx \int dx^\prime
\Tr[T^{I(0)}_{\mu\nu}(x,x^\prime)\sigma_{\alpha\beta}\gamma_5 ]
w^{\mu}_{\mu^\prime}\Tr[\sigma^{\alpha\beta}\gamma_5 \Phi^{\nu\mu^\prime}(x,x^\prime)]+\cdots\;,
\eee
in which other spin decompositions give higher twist contributions.

The numerators in the contraction $\Tr[T^{I(0)}_{\mu\nu}\sigma_{\alpha\beta}\gamma_5 ]$
can give terms proportional to $n_{\nu}n_{\mu}(q_{\alpha}n_{\beta}-n_{\alpha}q_{\beta})$,
$n_{\nu}d_{\perp,\mu\mu^{\prime\prime}}(q_{\alpha}n_{\beta}-n_{\alpha}q_{\beta})$,
and $n_{\nu}d_{\perp,\mu\mu^{\prime\prime}} \epsilon_{\perp,\alpha\beta}$.
The transversal tensors $d_{\perp,\alpha\beta}$ and $\epsilon_{\perp,\alpha\beta}$ are defined as 
$d_{\perp,\alpha\beta}=q_{\alpha}n_{\beta}+q_{\beta}n_{\alpha}-g_{\alpha\beta}$ and
 $\epsilon_{\perp,\alpha\beta}=\epsilon_{\alpha\beta\eta\lambda}q^{\eta}n^{\lambda}$.
The trace of $d_{\perp,\alpha\beta}$ is defined to be negative $d^{\alpha}_{\perp,\alpha}=-2$.
Since $\nu$ and $\mu$ indices in $\Phi_{M_2}^{\nu\mu}$ are antisymmetric under $\mu\leftrightarrow \nu$,
the terms proportional to $n_{\nu}n_{\mu}(q_{\alpha}n_{\beta}-n_{\alpha}q_{\beta})$ then vanish.
For those terms proportional to $n_{\nu}d_{\perp,\mu\mu^{\prime\prime}}(q_{\alpha}n_{\beta}-n_{\alpha}q_{\beta})$,
as they are contracted with $\Tr[\sigma^{\alpha\beta}\gamma_5 \Phi^{\nu\mu^\prime}(x,x^\prime)]$,
the $q_{\alpha}$ factor in $n_{\nu}d_{\perp,\mu\mu^{\prime\prime}}(q_{\alpha}n_{\beta}-n_{\alpha}q_{\beta})$ 
results in twist-4 contributions by using the property of the long distance propagator of the quark fields.
The terms proportional to 
$n_{\nu}d_{\perp,\mu\mu^{\prime\prime}} \epsilon_{\perp,\alpha\beta}$ lead to twist-3 contributions.
The final result appears as 
\bee
\int dx \int dx^\prime \frac{G_{\mu}^{\beta}(x,x^\prime)n^{\mu}n_{\beta}}{(x^\prime-x)x}\;,
\eee
where the function $G_{\mu}^{\beta}(x,x^\prime)$ is defined as
\bee
G_{\mu}^{\beta}(x,x^\prime)&=&\int\frac{d^4 l}{(2\pi)^4}\delta(x-l\cdot n)
\int\frac{d^4 k}{(2\pi)^4}\delta(x^\prime-x-k\cdot n)
\int d^4y \int d^4z \nn
&&\times e^{il\cdot z}e^{ik\cdot y}
\langle M_2|\bar{q}_2(z)\sigma_{\mu\alpha}\gamma_5 w^{\alpha}_{\alpha^\prime}
gG^{\beta\alpha^\prime}(y)q_3(0)|0\rangle\ .
\eee
Note that we have used the $G$-parity symmetry $x\leftrightarrow \bar{x}^\prime$ to simplify the above result.
This assumption is valid for $\pi$ mesons, but may not be appropriate for the $K$ or $\eta$ mesons. 
Therefore, it is noted that, in the above result, there exist symmetry breaking effects for $K$ and $\eta$ mesons. 
However, we will ignore such a corrections from the symmetry breaking in the following calculations. 
By referring to the definition \cite{Ball:1998je}
\bee
&&\langle M_2|\bar{q}_2(z)\sigma_{\mu\nu}\gamma_5 gG_{\alpha\beta}(y)q_3(0)|0\rangle\nn
&=&-i\frac{f_{M_2} m_{M_2}^2}{m_{q_2}+m_{\bar{q}_3}}
(q_{\alpha}q_\mu d_{\perp,\nu\beta}-q_{\alpha}q_{\nu}d_{\perp,\mu\beta}
-q_{\beta}q_{\mu}d_{\perp,\nu\alpha}+q_{\beta}q_{\nu}d_{\perp,\alpha\mu})
T(z,y)+\cdots\ ,
\eee
where 
\bee
T(z,y)=\int_0^1 dx \int_0^{\bar{x}} dx^\prime e^{-ixq\cdot z}e^{-i(x-x^\prime)q\cdot y}T(x,x^\prime)\ ,
\eee
we can arrive at the result 
\bee
&&\int dx \int dx^\prime \frac{G_{\mu}^{\beta}(x,x^\prime)n^{\mu}n_{\beta}}
{(x^\prime-x)x}\nn
&=&-\frac{2if_{M_2} m_{M_2}^2}{m_{q_2}+m_{\bar{q_3}}}\int dx \int dx^\prime \frac{T(x, x^\prime)}{(x^\prime-x)x}\ .
\eee

By using the normalization for $\langle M_1|\bar{q}_1(0)(1-\gamma_5)b(0)|\bar{B}\rangle$,
it is easy to derive the tree level three parton contributions for operator $ O_6$ as
\bee
\langle O_6\rangle_{1-gluon}
=\frac{2 A_{M_2}^{G3} m_{M_2}^2}{m_b(m_{q_2}+m_{\bar{q}_3})}
\langle O_1\rangle_f
\eee
with
\bee\label{AM2G3}
A_{M_2}^{G3}=2\int_0^1 dx \int_0^{\bar{x}} dx^\prime \frac{T_{M_2}(x^\prime,x)}{(x^\prime-x)x}\ .
\eee

We now explain the expansion with the covariant gauge $\partial\cdot A=0$. 
We first decompose $A^{\mu}(y)$ into its longitudinal and transversal components as
$A^{\mu}(y)=n\cdot A(y)q^{\mu}+ d^{\mu}_{\perp,\mu^\prime}A^{\mu^\prime}(y) $. 
The transversal part $d^{\mu}_{\perp,\mu^\prime}A^{\mu^\prime}(y)$ results in contributions of higher than twist-3 . 
The longitudinal part $n\cdot A(y)q^{\alpha}$ gives twist-3 contributions. 
Similar to the light-cone gauge, we need to transform the gluon fields into its field strength.
Here, it needs one transversal momentum $k_\perp$ factor from expansion of the parton amplitude
$T^{I(0)}_{\mu}(l,k)$ in Eq.~(\ref{twist3G-1}). 
The contraction of $T^{I(0)}_{\mu}(x,x^\prime)$ with $q^\mu n\cdot\Phi_{M_2}(x,x^\prime)$ leads to two parton 
gauge phase factor
\bee
\Tr[T^{I(0)}_{\mu}(x,x^\prime)q^\mu n\cdot\Phi_{M_2}(x,x^\prime)]
=\Tr[\frac{T^{I(0)}(x^\prime)-T^{I(0)}(x)}{x^{\prime }-x}n\cdot\Phi_{M_2}(x,x^\prime)]
\eee 
It is convenient to write 
$(k-\hat{k})^{\rho}=d^{\rho}_{\perp\rho^\prime}(k-\hat{k})^{\rho^\prime}+q\cdot k n^{\rho}$. 
Only transversal part 
$k^{\rho}_{\perp}=d^{\rho}_{\perp\rho^\prime}(k-\hat{k})^{\rho^\prime}$ contributes at twist-3. 
This is because the term $\partial T^{I(0)}_{\mu}/\partial k^{\nu}$ 
can have terms proportional to $g_{\mu\nu}$ and $\sigma_{\mu\nu}$. 
The terms related to $q\cdot k n^{\nu}$ leads to twist-4 contributions. 

For the transversal part $k_\perp$,
only $\sigma_{\mu\nu}$ terms can contribute. 
Let the $k^{\rho}_{\perp}$ factor absorbed into $\Phi^{\mu}(l,k)$ 
and use the replacement $k^{\nu}_{\perp}A^{\mu}(y)\to -i G^{\nu\mu}(y)$, 
we can derive the result
\bee\label{twist-3Cov}
\langle O_6\rangle^{t=3}_{1-gluon}&=&-2\int dx \int dx^\prime 
\Tr[\frac{\partial T^{I(0)}_{\mu}(x,x^\prime)}{\partial k^{\nu}}G^{\nu\mu}(x,x^\prime)]\nn
&&\times\langle M_1|\bar{q}_1(0)(1-\gamma_5)b(0)|\bar{B}\rangle\ ,
\eee
where 
\bee
\frac{\partial T^{I(0)}_{\mu}(x^\prime,x)}{\partial k^{\nu}}
=\frac{-i\sigma_{\mu\nu}}{(x^\prime-x)x q^2}(1+\gamma_5)
\eee
and 
\bee
G^{\nu\mu}(x,x^\prime)&=&\int\frac{d^4 l}{(2\pi)^4}\delta(x-l\cdot n)
\int\frac{d^4 k}{(2\pi)^4}\delta(x^\prime-x-k\cdot n)
\int d^4y \int d^4z \nn
&&\times e^{il\cdot z}e^{ik\cdot y}
\langle M_2|\bar{q}_2(z)igG^{\nu\mu}(y)q_3(0)|0\rangle\ .
\eee
The contraction of $\sigma_{\mu\nu}$ with $G^{\nu\mu}(x,x^\prime)$ gives
\bee
\Tr[i\sigma_{\mu\nu}G^{\nu\mu}(x,x^\prime)]=\frac{-2if_{M_2}m_{M_2}^2 q^2}{(m_{q_2}+m_{\bar{q}_3})}T(x,x^\prime)\ .
\eee
Note that the $q^2$ factor in the denominator of ${\partial T^{I(0)}_{\mu}(x^\prime,x)}/{\partial k^{\nu}}$ 
is cancelled by the $q^2$ factor in the numerator of $\Tr[i\sigma_{\nu\mu}G^{\nu\mu}(x,x^\prime)]$.
It is easy to see that Eq.~(\ref{twist-3Cov}) is equal to the result derived from the light-cone gauge. 
This explicitly shows the gauge invariance of the three parton contributions.

There are related diagrams, such as those in  Fig.~\ref{fig:fig3}(c) and \ref{fig:fig3}(d).
Because the spectator quark of the $\bar{B}$ meson can carry only a soft momentum,
this makes the relevant contributions associated with Fig.~\ref{fig:fig3}(c) and \ref{fig:fig3}(d) 
dominated by soft gluons as the form factors $F_{+,0}^{B\to M_1}$. 
In addition, the relevant contributions are of $O(m_b^{-2})$ with respect to 
the leading twist amplitude. 
It can be understood as following.
The sum of the lower parts of the diagarms in Fig.~\ref{fig:fig3}(c) and \ref{fig:fig3}(d) is proportional to
\bee
\langle M_1|\bar{q}_1
\left[\frac{2p_{\nu}+\gamma_{\nu}\s{k}}{2p\cdot k}\Gamma_i-
\Gamma_i\frac{2P_{b\nu}- \s{k}\gamma_{\nu}}{2P_b\cdot k}\right]b|\bar{B}\rangle\;,
\eee
where $k$ is the momentum of the gluon from $M_2$ and the equation of motions for $b$ and $q_1$ quarks have been used. 
After taking the collinear limit, $k\to x^{\prime}q$,
we write the expression  as
\bee
A q_{\nu} + B_{\mu\nu}q^{\mu}\;,
\eee
where
\bee
A &\equiv& \frac{1}{x^{\prime} Q^2}\langle M_1|\bar{q}_1\Gamma_i|\bar{B}\rangle\;,\\
B_{\mu\nu} &\equiv& \frac{1}{2 x^{\prime} Q^2}\langle M_1|\bar{q}_1 
(\gamma_{\nu}\gamma_{\mu}\Gamma_i + \Gamma_i \gamma_{\mu}\gamma_{\nu})b|\bar{B}\rangle\;.
\eee
For light-cone gauge, only $A$ term contributes.
As for the covariant gauge, only $B_{\mu\nu}$ term contributes.
The only contributions come from $(V-A)(V\pm A)$ operators.
This implies that the upper parts of the diagarms in Fig.~\ref{fig:fig3}(c) and \ref{fig:fig3}(d) are proportional to
the twist-4 LCDA of $M_2$.
The combination of the upper and the lower parts gives a $O(m_b^{-2})$ contributions with respect to 
the leading twist amplitude.

There are possibilities that the additional gluon of the $M_2$ meson can interact with the spectator quark of 
the $\bar{B}$ meson.
Since the spectator quark carries a soft momentum,
the momentum conservation at the interaction vertex prevents the momentum of the gluon from
being collinear to the $M_2$ meson's momentum.
Therefore, there require additional radiative gluons interacting between the other parton lines 
and the spectator quark line to
make the momentum of the gluon to be collinear to the $M_2$ meson's momentum.
This results in contributions of order $O(\alpha_s)$.
We identify the relevant contributions as $O(\alpha_s)$ three parton corrections.
As mentioned previously, we plan to discuss these contributions in other places \cite{Yeh:2007}.

The total twist-3 contribution from operator $O_6$ is then equal to
\bee
\langle O_6\rangle^{t=3}=\frac{2(1+A_{M_2}^{G3}) m_{M_2}^2}{m_b(m_{q_2}+m_{\bar{q}_3})}\langle O_1\rangle_f\ .
\eee
For operator $O_8$, there are similar results.

\section{Applications}
For penguin dominant $B\to\pi K$ decays,
the relevant decay amplitudes under QCD factorization are parametrized as the following \cite{Beneke:2001ev}
\bee
&&A(B^{-}\to \pi^{-}\bar{K}^0)
=\lambda_{p}\left[(a_4^p -\frac{1}{2} a_{10}^p)+r_{\chi}^{K}(a_6^p -\frac{1}{2} a_8^p)\right] A_{\pi K}\nn
&&\hspace{3.5 cm}+ (\lambda_u b_2 + (\lambda_u +\lambda_c)(b_3 + b_3^{EW}))B_{\pi K}\;,\nn
&& - \sqrt{2}A(B^-\to \pi^0 K^-)=[\lambda_u a_1 +\lambda_p ( a_4^p + a_{10}^p)
+ \lambda_p r_{\chi}^{K}(a_6^p + a_8^p) ]A_{\pi K} \nn
&&\hspace{4.5 cm}+ [\lambda_u a_2 + \lambda_p \frac{3}{2}(-a_7 + a_9)]A_{K\pi}\nn
&&\hspace{4.5 cm}+ (\lambda_u b_2 + (\lambda_u \lambda_c)(b_3 + b_3^{EW}))B_{\pi K}\;,\nn
&& -A(\bar{B}^{0}\to\pi^+ K^{-})=[\lambda_u a_1 + \lambda_p (a_4^p + a_{10}^p) 
+\lambda_p r_{\chi}^K (a_6^p + a_8^p)]A_{\pi K}\nn
&&\hspace{3.5 cm}+((\lambda_u +\lambda_c )( b_3 - \frac{1}{2}b_3^{EW}))B_{\pi K}\;\nn
&&\sqrt{2}A(\bar{B}^0 \to \pi \bar{K}^0)=A(B^- \to \pi^- \bar{K}^0) + \sqrt{2}A(B^-\to\pi^0 K^-)
-A(\bar{B}^0\to \pi^+ K^-)
\eee
where $\lambda_p =V_{pb}V_{ps}^*$, $a_i\equiv a_i (\pi K)$, 
and $\lambda_p a_i^p=\lambda_u a^u_i + \lambda_c a^c_i$.
The $CP$ conjugation of decay amplitudes are obtained by replacing $\lambda_p \to \lambda_p^*$ for the above amplitudes.
The factorized matrix elements are defined as
\bee
A_{\pi K}&=&i\frac{G_F}{\sqrt{2}}(m_B^2-m_{\pi}^2)F^{B\to\pi}_0 (m_K^2 )f_K\;,\nn
A_{K \pi }&=&i\frac{G_F}{\sqrt{2}}(m_B^2-m_{K}^2)F^{B\to K}_0 (m_{\pi}^2 )f_{\pi}\;.
\eee
The form factors are defined
\bee
\langle P(p)|\bar{q}\gamma^{\mu}b|\bar{B}\rangle
=F_+^{B\to P}(q^2)(P_{B}^{\mu} + p^{\mu})
+[F_0^{B\to P}(q^2)-F_+^{B\to P}(q^2)]\frac{m_B^2-m_P^2}{q^2}q^{\mu}\;.
\eee
The form factors coincide as $q^2=0$, $F^{B\to P}_+ (0)=F_0^{B\to P}(0)$.
The expressions for the parameters $a_i$ are referred to \cite{Beneke:2000ry,Beneke:2001ev}.
For numerical calculations, we will use the following input parameters
\bee
\begin{array}{cccc}
\Lambda_{\bar{MS}}^{(5)}=0.225 \text{GeV}\ , & m_b(m_b)=4.2 \text{GeV} & m_c(m_b)=1.3 \text{GeV}\ , & m_s(2\text{GeV})=0.090\text{GeV}\ , \\
|V_{cb}|=0.41 \ , & |V_{ub}/V_{cb}|=0.09 \ ,& \gamma=70^{\circ} \ ,& \tau(B^{-})=1.67(\text{ps}) \ , \\
\tau(B_d)=1.54(\text{ps}) \ , & f_{\pi}=131\text{MeV} \ ,& f_K = 160\text{MeV}\ , & f_B=200\text{MeV}\ , \\
F_0^{B\to\pi}=0.28 \ ,& F_0^{B\to K}=0.34 \ .& &\\
\end{array}
\eee
For $\lambda_u$ and $\lambda_c$, we take the following convention for their parametrization
\bee
\frac{\lambda_u}{\lambda_c}=\tan^2\theta_c R_b e^{-i\gamma}
\eee
where
\bee
&& \tan^2\theta_c =\frac{\lambda^2}{1-\lambda^2}\;,\nn
&& R_b =\frac{1-\lambda^2/2}{\lambda}|\frac{V_{ub}}{V_{cb}}|\;,\nn
&& \lambda = |V_{us}|\;.
\eee
The value of $\lambda$ is taken as $0.22$.

By using previous input parameters, 
we list the values of $a_i$, $i=1,\cdots, 10$, 
and $b_j$, $j=1,\cdots, 3$, calculated at the scale $m_b=4.2 \text{GeV}$ as below
\bee 
\begin{array}{lll}
a_1 = 0.995+0.018 i\ , & a_2 = 0.209-0.104 i\ , & a_3 = -0.003+0.003 i \ ,\\
a_4^u = -0.031-0.013 i & a_4^c =  -0.030 + 0.027 i\ , & a_5 = 0.007 - 0.004 i \ , \\
r_{\chi}^K a^u_6 = -0.050 - 0.015 i \ , & r_{\chi}^K a^c_6 = -0.047 - 0.005 i \ , & a_7/\alpha = 0.007 + 0.006 i \ ,\\
r_{\chi}^K a^u_8/\alpha = 0.087 -0.043 i \ , & r_{\chi}^K a^c_8/\alpha = 0.094 - 0.021 i \ , & a_9/\alpha = -1.135 - 0.024 i \ ,\\
a^u_{10}/\alpha = -0.175 + 0.093 i \ , & a^c_{10}/\alpha = -0.175 + 0.093 i \ , & r_A b_1=0.021\ ,\\
r_A b_2 =-0.008 \ ,& r_A b_3 =-0.006 \ , & r_A b_3^{EW}/\alpha = -0.018 \ ,  
\end{array}
\eee
in which
\bee
r_A = \frac{B_{\pi K}}{A_{\pi K}} = \frac{f_B f_{\pi}}{m_B^2 F^{B\to\pi}_0 (0)}\; ,
\eee
and 
\bee
r_{\chi}^{K}=\frac{2 m_K^2}{m_b(m_{q}+ m_s )}\;.
\eee
The tree level three parton contributions modify the parameters $r_{\chi}^K $ as $r_{\chi}^K (1+A_{M_2}^{G_3})$
with parameter $A_{M_2}^{G_3}$ defined in Eq.~(\ref{AM2G3}). 
The value of $A_{M_2}^{G_3}$ depends on the model of 
the three parton distribution amplitude $T_{M_{2}}(x,x^\prime)$. 
Here we employ the model derived from the light-cone sum rule \cite{Ball:1998je}
\bee
T(x,x^{\prime})=360\eta x x^{\prime}(x-x^\prime)^2
(1+\frac{\omega}{2}(7(x-x^\prime )-3)),
\eee
where the parameters are assumed to be $\eta=0.015$ and $\omega=-3.0$ for $M_2=\pi, K,$ or $\eta$.
This give us $A_{M_2}^{G_3}=0.585$. 

The branching ratio for a $\bar{B}\to \pi K$ decay is given by this expressions
\bee
Br(\bar{B}\to \pi K)=\frac{\tau_B}{16\pi m_B}|A(\bar{B}\to\pi K)|^2\;.
\eee
We can use the above formula to predict $CP$ averaged branching ratios for $B\to\pi K$ decays.
The predictions with three parton corrections  in units of $10^{-6}$ are given as
\bee\label{three-parton-collinear}
&& Br(B^{-}\to\pi^{-} \bar{K}^{0})=19.0\;,\nn
&& Br(B^{-}\to\pi^{0} K^{-})=10.0\;,\nn
&& Br(\bar{B}^{0}\to\pi^{+} K^{-})=16.1\;,\nn
&& Br(\bar{B^0}\to\pi^{0} \bar{K^{0}})=7.7\;.
\eee
For comparison, we also list the predictions with only two parton contributions in units of $10^{-6}$ in the following,
\bee\label{two-parton-collinear}
&& Br(B^{-}\to\pi^{-} \bar{K}^{0})=11.2\;,\nn
&& Br(B^{-}\to\pi^{0} K^{-})=6.1\;,\nn
&& Br(\bar{B}^{0}\to\pi^{+} K^{-})=9.4\;,\nn
&& Br(\bar{B^0}\to\pi^{0} \bar{K^{0}})=4.4\;.
\eee
For reference, we enlist the experimental data in units of $10^{-6}$ summarized by the HFAG group 
\cite{Barberio:2006bi} 
\bee\label{experiments}
&& Br(B^{-}\to\pi^{-} \bar{K}^{0})=23.1\pm 1.0\;,\nn
&& Br(B^{-}\to\pi^{0} K^{-})=12.8\pm 0.6\;\nn
&& Br(\bar{B}^{0}\to\pi^{+} K^{-})=19.7\pm 0.6\;\nn
&& Br(\bar{B^0}\to\pi^{0} \bar{K^{0}})=10.0\pm 0.6\;.
\eee
By comparing the predictions with or without the three parton corrections,
one may notice that the the predicted branching ratios are significantly
enhanced by about $1.65\sim 1.75 $ times in their magnitudes.

The two parton predictions for the $\bar{B}\to \pi K$ decays made here are consistent with the findings of 
previous studies using QCD factorization approach \cite{Du:2001hr,Beneke:2001ev}.
The two parton predictions made in \cite{Du:2001hr} are much lower than the experimental data under the QCD factorization approach.
Because the calculations of the three parton corrections were inaccessible in their studies,
this led them to conclude that the QCD factorization is impossible to explain the penguin dominant 
$\bar{B}\to \pi K$ decays.
The two parton predictions made in \cite{Beneke:2001ev} for $\bar{B}\to\pi K$ decays are still lower than the data.
Only extending the predictions by using extreme limits of input parameters can make the predictions to 
be consistent with the measurements.
This seems not a reasonable solution from the theoretical point of view.
On the other hand, as shown in the above,
our approach has shown that
the predictions with three parton contributions are more close to the data than the two parton predictions.
Although the predictions with three parton contributions are still lower than the experimental data,
the $O(\alpha_s)$ corrections can improve the predictions.

\section{discussions and conclusions}
The significance of the three parton contributions for penguin dominant $\bar{B}\to \pi K$ decays can also be seen from
a phenomenological point of view.
By appropriate arrangement, the parameters $a_i$ can be calculated for $B\to\pi\pi$ decays.
For the pure penguin $B^{-}\to \pi^- \bar{K}^0$,
its dominant contributions arise from the $|a_4^{c}(\pi K)+r_{\chi}^K a_6^{c}(\pi K)|$ term.
The uncertainty due to the form factors can be eliminated by considering the ratio
between the decay rates of $B^-\to \pi^- \bar{K}^0$ and $B^-\to \pi^-\pi^0$ as the following
\bee
\left|\frac{a_4^{c}(\pi K)+r_{\chi}^K a_6^{c}(\pi K)}{a_1(\pi\pi)+a_2(\pi\pi)}\right|
&=&\frac{|V_{ub}|}{|V_{cb}|}\frac{f_{\pi}}{f_K}
\left[\frac{\Gamma(B^-\to \pi^- \bar{K}^0)}{2\Gamma(B^-\to \pi^-\pi^0)}\right]^{1/2}\nn
&=& 0.105\pm 0.001\;,
\eee
where the error comes from the branching ratios.
In the above, we have used the branching ratio $Br(B^-\to \pi^-\pi^0)=5.7\pm 0.4$. 
According to QCD factorization calculations, $|a_1(\pi\pi)+a_2(\pi\pi)|=1.17$ and
$|a_4^{c}(\pi K)+r_{\chi}^K a_6^{c}(\pi K)|=0.08$ with only two parton contributions.
This gives the prediction of the ratio to be $0.066$, which is lower than the experimental value.
By adding the tree level three parton contributions, the factor $r_{\chi}^K$ then becomes $r_{\chi}^K(1+A_{K}^{G3})$
and makes $|a_4^{c}(\pi K)+r_{\chi}^K(1+A_{K}^{G3}) a_6^{c}(\pi K)|=0.104$.
The predicted ratio is changed to be $0.089$, which is closer to the two parton prediction.
The above fact may also indicate that the three parton corrections could be important for 
understanding the penguin dominant $\bar{B}\to\pi K$ decays under the QCD factorization approach.

In order to make sure that the three parton contributions are indeed significant 
and also compatible with the QCD factorization,
an important task is to finish $O(\alpha_s)$ calculations for the three parton contributions.
The related work in this direction has been proceeded and will be reported in our another preparing paper 
\cite{Yeh:2007}. 

\begin{figure*}
\includegraphics{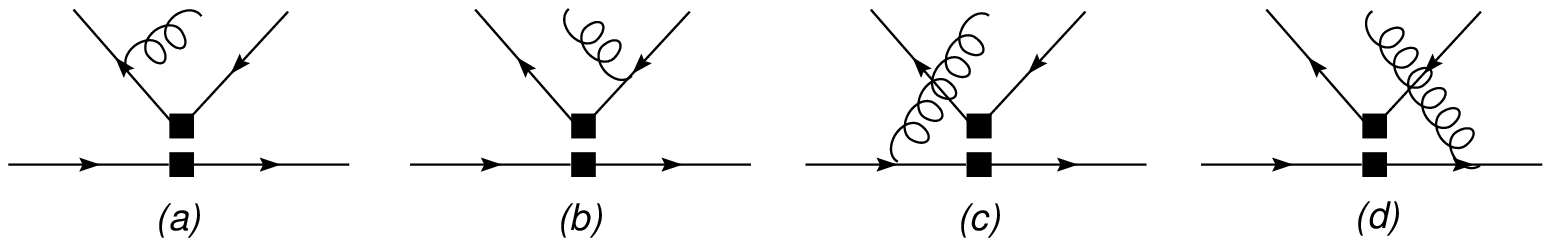}
\caption{\label{fig:fig3}The Feynman diagrams describe the tree level five parton amplitude, $T^{I(0)}_{ij,\mu}$. 
The square symbol represents the vertex of weak interactions.}
\end{figure*}

\begin{figure*}
\includegraphics{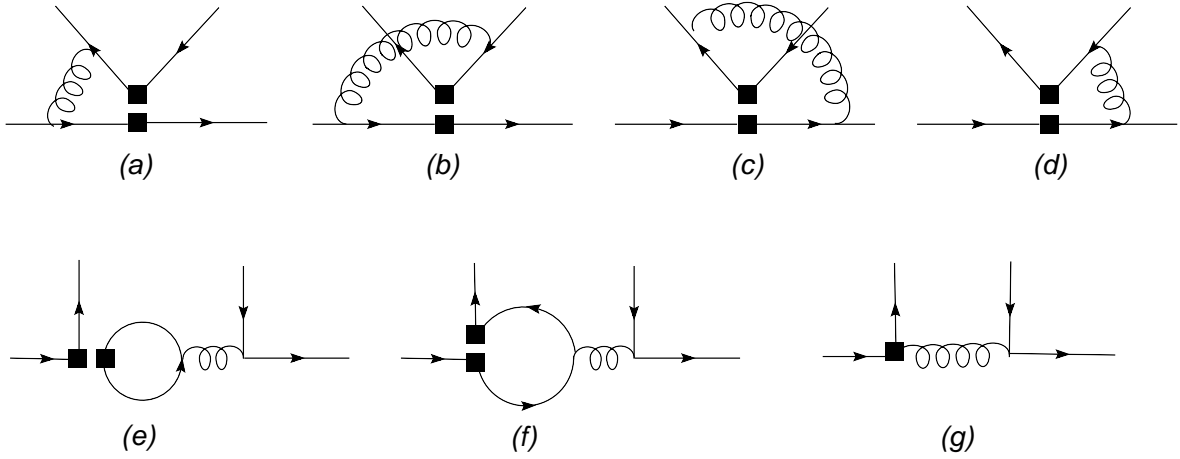}
\caption{\label{fig:fig4}The Feynman diagrams describe the $O(\alpha_s)$ four parton amplitude, $T^{I(1)}_{ij}$. 
The square symbol represents the vertex of weak interactions.}
\end{figure*}

\begin{figure*}
\includegraphics{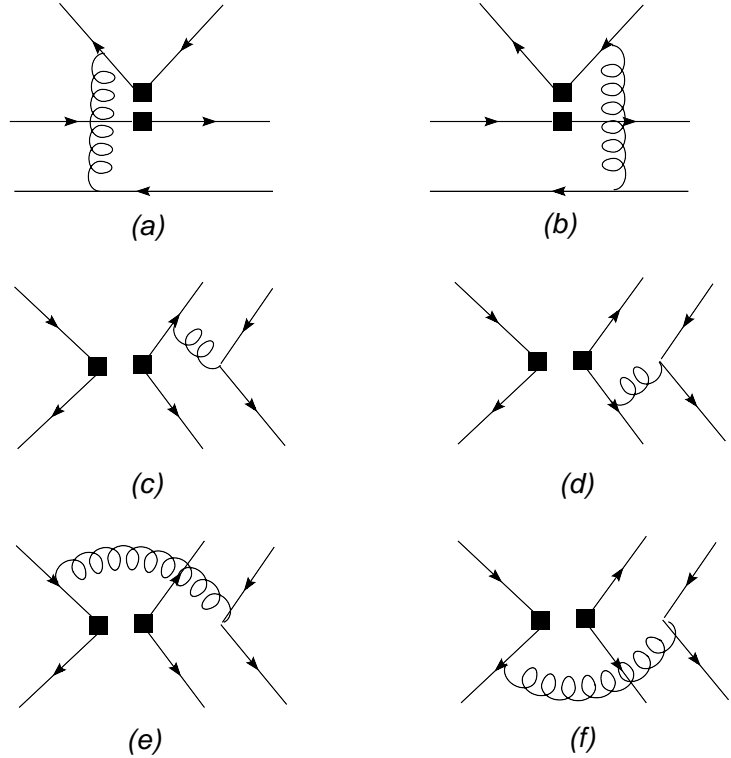}
\caption{\label{fig:fig5}The Feynman diagrams for the $O(\alpha_s)$ six parton amplitude, $T^{II(1)}$. 
The square symbol represents the vertex of weak interactions.}
\end{figure*}

\begin{figure*}
\includegraphics{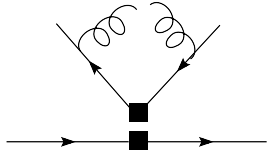}
\caption{\label{fig:fig6}The Feynman diagram for the six parton amplitude with $|q\bar{q}gg\rangle$ Fock state. 
The square symbol represents the vertex of weak interactions.}
\end{figure*}

There are similar three parton contributions having been calculated under the light-cone sum rule 
\cite{Khodjamirian:2000mi, Yang:2003sg}. 
In \cite{Khodjamirian:2000mi}, 
the twist-3 three parton contributions associated with soft gluons are calculated 
in the framework of light-cone sum rule.
The contributions are shown negligible in $\bar{B}\to \pi\pi$ decays.
In \cite{Yang:2003sg}, the contributions from the diagrams similar to Fig.~\ref{fig:fig3}(c) and \ref{fig:fig3}(d)
were calculated for $B\to \pi\omega $ decays.
They are found to be twist-4 and vanishing.
In addition, significant effects were found due to the three parton Fock state of the $\pi$ 
in the $B\to \pi\omega $ decays.
Since they are dominated by soft gluons, it is better determined by QCD sum rule.
As shown in \cite{Yang:2003sg}, in the Euclidean region of $(p + q)^2$,
the relevant contributions are from the twist-3 and twist-4 three parton LCDAs of the $\pi$.
As mentioned before, we identify these power corrections as non-partonic ones.
From the theoretical point of view,
we suggest that the partonic and non-partonic power corrections should be distinquished under the QCD factorization,
although they may be equally important in phenomenology.  

For comparison, we employ the replacing rules for the $a_i$ coefficients 
\cite{Yang:2003sg} to account for the three parton effects from the $M_1$ meson.
The rule is
\bee
a_{2i}&\to& a_{2i}+[1+(-1)^{\delta_{3i}+\delta_{4i}}]C_{2i-1}f_{3}/2\;,\nn
a_{2i-1}&\to& a_{2i-1}+(-1)^{\delta_{3i}+\delta_{4i}}C_{2i}f_{3}\;,
\eee
where $i=1,\cdots,5$, 
and $C_i$ are the Wilson coefficients calculated at the scale $\mu_{h}=1.45$ GeV,
and $f_3=0.12$,
which is assumed to be universal. 
With these three parton corrections, the predicted $CP$ averaged branching ratios for $\bar{B}\to \pi K$ 
in units of $10^{-6}$ are
\bee
&& Br(B^{-}\to\pi^{-} \bar{K}^{0})=9.5\;,\nn
&& Br(B^{-}\to\pi^{0} K^{-})=5.9\;,\nn
&& Br(\bar{B}^{0}\to\pi^{+} K^{-})=8.4\;,\nn
&& Br(\bar{B^0}\to\pi^{0} \bar{K^{0}})=3.4\;,
\eee 
which becomes smaller than those predictions in Eq.~(\ref{two-parton-collinear}).
\vspace{0.5cm}
\noindent
\begin{center}
{\bf Acknowledgments}
\end{center}
The author would like to appreciate the referees for giving this paper many helpful suggestions.
This work was supported in part by the National
Science Council of R.O.C. under Grant Numbers NSC92-2112-M-142-001, NSC93-2112-M-142-001 and NSC95-2112-M-142-001.

\newpage
\newpage
\newpage
\newpage
\newpage
\end{document}